\documentclass[aps,pra,10pt,twocolumn,notitlepage,floatfix,superscriptaddress,longbibliography]{revtex4-1} 

\usepackage[dvips]{graphicx}
\usepackage{subfigure}
\usepackage{amsmath}
\usepackage{amssymb}
\usepackage{bm}
\usepackage{color}
\usepackage[colorlinks=true,citecolor=blue,urlcolor=blue,linkcolor=blue,hyperfigures=true]{hyperref}
\usepackage{multirow}
\usepackage{comment}

\graphicspath{{./}}
\def\Eq{Eq.~}
\def\Eqs{Eqs.~}
\def\Fig{Fig.~}
\def\Figs{Figs.~}

\def\Sec{Sec.~}

\def\be{\begin{equation}}
\def\ee{\end{equation}}
\def\bea{\begin{eqnarray}}
\def\eea{\end{eqnarray}}

\def\ie{\textit{i.e.}~}
\def\eg{\textit{e.g.}~}

\def\etal{\textit{et al}~}
\def\dd{\mathrm{d}}

\def\x10{\times 10}
\def\half{\frac{1}{2}}
\newcommand{\bra}[1]{\left\langle #1 \right|}
\newcommand{\ket}[1]{\left| #1 \right\rangle}

\begin{document}

\title{Multi-Dimensional Atom Optics and Interferometry}

\author{B. Barrett}
\affiliation{iXblue, 34 rue de la Croix de Fer, 78105 Saint-Germain-en-Laye, France}
\affiliation{LP2N, Laboratoire Photonique, Num\'{e}rique et Nanosciences, Universit\'{e} Bordeaux--IOGS--CNRS:UMR 5298, 1 rue Fran\c{c}ois Mitterrand, 33400 Talence, France}
\author{P. Cheiney}
\affiliation{iXblue, 34 rue de la Croix de Fer, 78105 Saint-Germain-en-Laye, France}
\affiliation{LP2N, Laboratoire Photonique, Num\'{e}rique et Nanosciences, Universit\'{e} Bordeaux--IOGS--CNRS:UMR 5298, 1 rue Fran\c{c}ois Mitterrand, 33400 Talence, France}
\author{B. Battelier}
\affiliation{LP2N, Laboratoire Photonique, Num\'{e}rique et Nanosciences, Universit\'{e} Bordeaux--IOGS--CNRS:UMR 5298, 1 rue Fran\c{c}ois Mitterrand, 33400 Talence, France}
\author{F. Napolitano}
\affiliation{iXblue, 34 rue de la Croix de Fer, 78105 Saint-Germain-en-Laye, France}
\author{P. Bouyer}
\affiliation{LP2N, Laboratoire Photonique, Num\'{e}rique et Nanosciences, Universit\'{e} Bordeaux--IOGS--CNRS:UMR 5298, 1 rue Fran\c{c}ois Mitterrand, 33400 Talence, France}

\date{\today}


\begin{abstract}
We propose new multi-dimensional atom optics that can create coherent superpositions of atomic wavepackets along three spatial directions. These tools can be used to generate light-pulse atom interferometers that are simultaneously sensitive to the three components of acceleration and rotation, and we discuss how to isolate these inertial components in a single experimental shot. We also present a new type of atomic gyroscope that is insensitive to parasitic accelerations and initial velocities. The ability to measure the full acceleration and rotation vectors with a compact, high-precision, low-bias inertial sensor could strongly impact the fields of inertial navigation, gravity gradiometry, and gyroscopy.
\end{abstract}

\maketitle


Inertial sensors based on cold atoms and light-pulse interferometry \cite{Borde1989, Kasevich1991, Cronin2009} exhibit exquisite sensitivity that could potentially revolutionize a variety of fields including geophysics and geodesy \cite{Carraz2014, Menoret2018}, gravitational wave detection \cite{Canuel2018}, tests of fundamental laws and inertial navigation \cite{Hogan2009, Barrett2014a}. Their state-of-the-art sensitivity and ultra-low measurement bias are particularly appropriate for long-term integration as in precision measurements \cite{Bouchendira2011, Rosi2014} or space experiments \cite{Becker2018}. They also offer great potential for autonomous inertial navigation systems \cite{Jekeli2005, Battelier2016, Fang2016, Cheiney2018}, where the attitude and position of a moving body is determined by integrating the equations of motion.

The measurement principle of light-pulse atom interferometers (AIs) is linked to a retro-reflected laser beam that is referenced to an atomic transition. This defines a phase ruler to which the free-falling atom's trajectory is compared \cite{Borde2001}, in analogy to classical falling-corner-cube gravimeters \cite{Niebauer1995}. In general, the direction of the retro-reflected beam defines the inertially-sensitive axis of these quantum sensors. They can be sensitive to accelerations \cite{Peters1999, Gillot2014, Freier2016, Hardman2016, Bidel2018} and acceleration gradients \cite{Snadden1998, Rosi2015, Asenbaum2017} parallel to the effective optical wavevector $\bm{k}$, and to rotations perpendicular to the plane defined by $\bm{k} \times \bm{v}_0$ \cite{Gustavson1997, Stockton2011, Barrett2014b, Tackmann2014, Rakholia2014, Dutta2016, Yao2018}, where $\bm{v}_0$ is the initial velocity of the atomic source. So far, the challenge of realizing multi-axis inertial measurements has been addressed only in a sequential manner \cite{Canuel2006, Wu2017}, where the direction of $\bm{k}$ was changed between measurement cycles. In this work, we propose new multi-dimensional AI geometries that are \emph{simultaneously} sensitive to accelerations and rotations in 3D, and can discern their vector components within a single shot.

In what follows, we define a multi-dimensional AI as one where the light interaction exchanges momentum with the atomic sample along more than one spatial direction at a time. This momentum exchange is accompanied by an independent phase shift along each axis, which is imprinted on the corresponding diffracted wavepacket \cite{Borde2004}. This mechanism creates a unique type of atom-optical element that satisfies all the requirements of a multi-dimensional AI---enabling one to split, reflect and recombine matter waves along two or more axes simultaneously.


\begin{figure}[!b]
  \centering
  \includegraphics[width=0.48\textwidth]{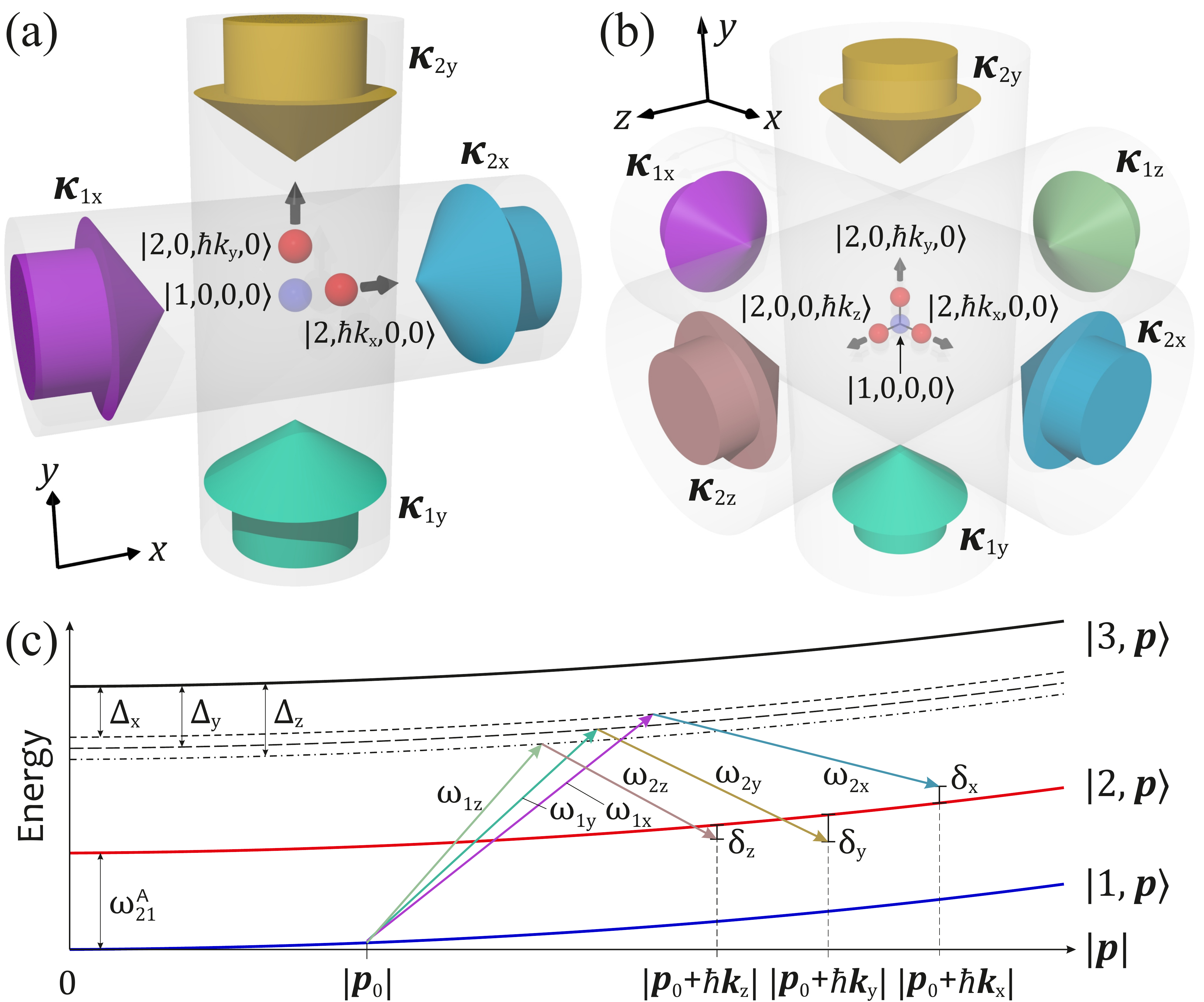}
  \caption{(a,b) Multi-dimensional atom-optical beamsplitters realized with mutually-orthogonal pairs of independent, counter-propagating Raman beams. (c) Energy-momentum diagram showing velocity-sensitive Raman transitions associated with each pair of beams shown in (a,b). To avoid excitation of parasitic resonances, different detunings $\Delta_{\mu}$ are used for each beam pair. This leads to a small difference in the wavevectors $k_{\mu}$, which has been exaggerated for clarity.}
  \label{fig:2D+3DAtomOptics}
\end{figure}

We have developed a semi-classical model for 3D atom optics involving Raman transitions \cite{SupMat}. Figure \ref{fig:2D+3DAtomOptics} shows examples of 2D and 3D atom-optical beamsplitters, where mutually-orthogonal pairs of counter-propagating Raman beams couple an atom with initial momentum $\bm{p}_0$ to two or three diffracted states moving perpendicular to one another. Along each axis $\mu = x,y,z$, two beams with wavevectors $\bm{\kappa}_{1\mu}$ and $\bm{\kappa}_{2\mu}$, and corresponding frequencies $\omega_{1\mu}$ and $\omega_{2\mu}$, excite two-photon Raman transitions \footnote{These multi-dimensional atom optics can also be achieved using Bragg transitions.} between two ground states $\ket{1}$ and $\ket{2}$ separated by frequency $\omega_{21}^{\rm A}$. During this process, a momentum $\hbar \bm{k}_{\mu} = \hbar (\bm{\kappa}_{1\mu} - \bm{\kappa}_{2\mu})$ is transferred to the atom, where $\hbar$ is the reduced Planck's constant and $\bm{k}_{\mu} \simeq 2\bm{\kappa}_{1\mu}$ is the effective Raman wavevector along axis $\mu$ \cite{Aspect1989, Moler1992}. The laser frequencies $\omega_{n\mu}$ are detuned by $\Delta_\mu$ from an intermediate excited state $\ket{3,\bm{p}_0}$ as shown in \Fig \ref{fig:2D+3DAtomOptics}(c), such that $|\Delta_\mu|$ is large compared to the natural linewidth of the atomic transition and, for $\mu \neq \nu$, $|\Delta_{\mu} - \Delta_{\nu}|$ is much larger than the effective Rabi frequency. This second condition strongly inhibits scattering processes involving absorption along one axis and re-emission along an orthogonal one. In the region of beam overlap, an atom initially in $\ket{1,\bm{p}_0}$ undergoes 3D diffraction---splitting the wavepacket into a superposition of this undiffracted state and three orthogonal diffracted states: $\ket{2, \bm{p}_0 + \hbar \bm{k}_{\mu}}$. The dynamics of this coherent 3D diffraction process are described by Rabi oscillations in an effective 4-level system: $\ket{\Psi} = \mathcal{C}_0 \ket{1,0,0,0} + \mathcal{C}_x \ket{2,\hbar k_x,0,0} + \mathcal{C}_y \ket{2,0,\hbar k_y,0} + \mathcal{C}_z \ket{2,0,0,\hbar k_z}$, where the states are labelled by their internal energy and the photon momentum transfer along each direction. This system exhibits Rabi oscillations in the population between states, where the vector of state amplitudes $\bm{\mathcal{C}}^{\rm T} = (\mathcal{C}_0,\mathcal{C}_x, \mathcal{C}_y, \mathcal{C}_z)$ evolves according to
\be
  \label{C(t)}
  \bm{\mathcal{C}}(t) = \exp\left[ -i \begin{pmatrix}
    0 & \chi_x^* & \chi_y^* & \chi_z^* \\
    \chi_x & -\delta_x & 0 & 0\\
    \chi_y & 0 & -\delta_y & 0\\
    \chi_z & 0 & 0 & -\delta_z
  \end{pmatrix} t \right] \bm{\mathcal{C}}(0).
\ee
Here, $\chi_{\mu}$ is the Rabi frequency and $\delta_{\mu} \simeq (\omega_{1\mu} - \omega_{2\mu}) - \omega_{21}^{\rm A} - \delta_{\mu}^{\rm D} - \delta_{\mu}^{\rm R}$ is the two-photon detuning of each beam pair, where $\delta_{\mu}^{\rm D} = \bm{k}_{\mu} \cdot \bm{v}_0$ is the Doppler shift for initial velocity $\bm{v}_0 = \bm{p}_0/m$, $\delta_{\mu}^{\rm R} = \hbar k_{\mu}^2/2m$ is a photon recoil shift, and $m$ is the atomic mass. For the special case when $\delta_x = \delta_y = \delta_z \equiv \delta$, the effective Rabi frequency for this system can be written analytically as $\Omega_{\rm Rabi} = \frac{1}{2} \sqrt{\delta^2 + 4(|\chi_x|^2 + |\chi_y|^2 + |\chi_z|^2)}$.

\begin{figure}[!t]
  \centering
  \includegraphics[width=0.48\textwidth]{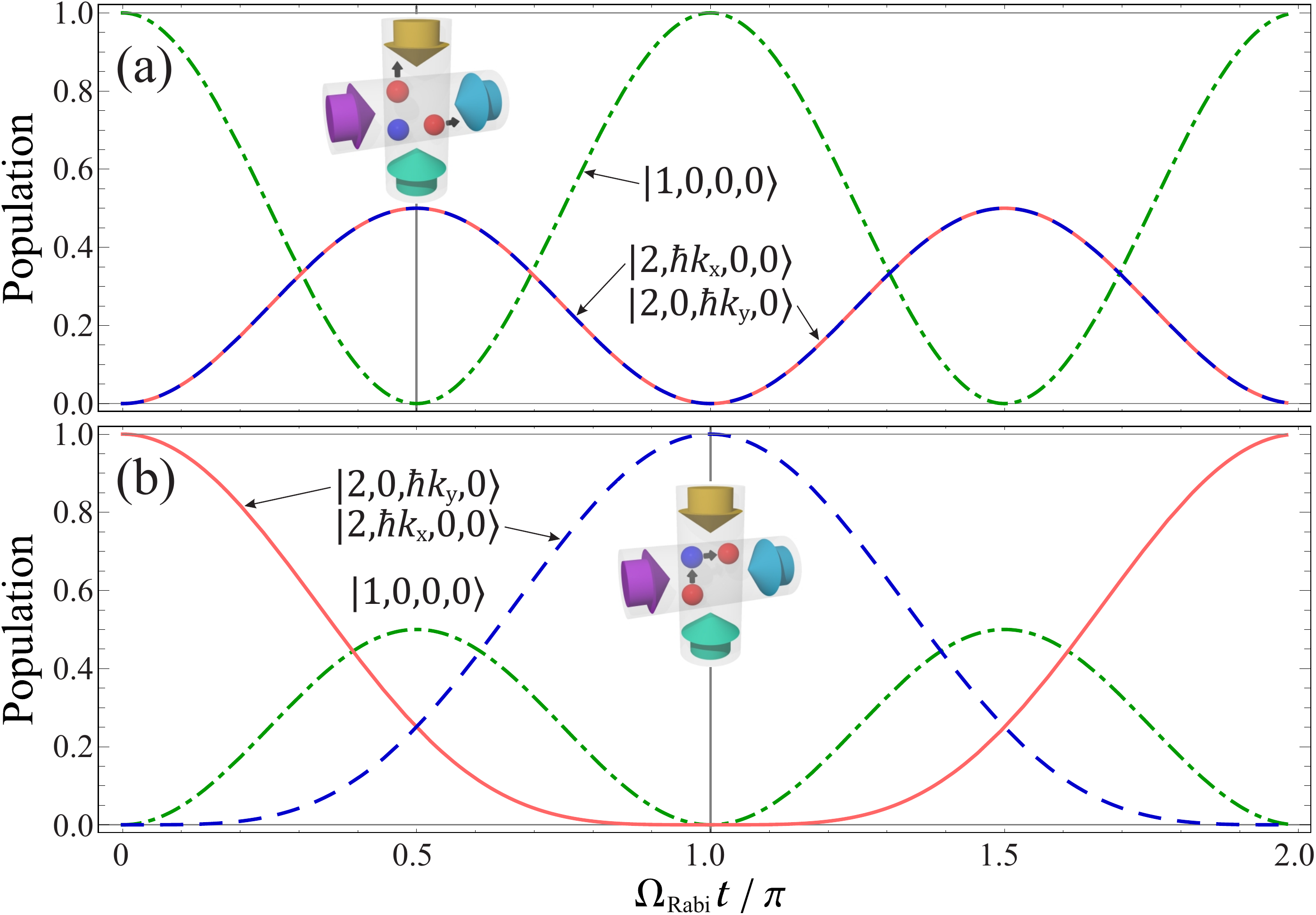}
  \caption{Rabi oscillations for a 2D beamsplitter (a) and mirror (b) realized using pulse areas of $\Omega_{\rm Rabi} t = \pi/2$ and $\pi$, respectively. The insets show the corresponding physical picture.}
  \label{fig:2DAtomOptics-Rabi}
\end{figure}

We now discuss the specific case of 2D atom optics. Figure \ref{fig:2DAtomOptics-Rabi}(a) displays Rabi oscillations corresponding to a 2D beamsplitter, where population is transferred between atoms initially in $\ket{1,0,0,0}$ and the two diffracted states $\ket{2,\hbar k_x,0,0}$ and $\ket{2,0,\hbar k_y,0}$. A beamsplitter is achieved at an interaction time $\tau$ corresponding to a pulse area $\Omega_{\rm Rabi}\tau = \pi/2$, where 50\% of the population is accumulated in the two target states. Contrary to a 1D beamsplitter, where one usually desires a 50/50 superposition of the initial and final states, here the population in the initial state is fully depleted---closely resembling 1D double-diffraction beamsplitters \cite{Leveque2009, Malossi2010, Giese2013, Kuber2016, Ahlers2016}. Similarly, \Fig \ref{fig:2DAtomOptics-Rabi}(b) shows the Rabi oscillation corresponding to a 2D mirror, where a $\pi$-pulse of duration $2\tau$ achieves 100\% population transfer between $\ket{2,\hbar k_x,0,0}$ and $\ket{2,0,\hbar k_y,0}$. The resonance frequency for this transition is identical to the population-reversed case ($\ket{2,0,\hbar k_y,0} \to \ket{2,\hbar k_x,0,0}$), which is ideal for reflecting the two arms of an interferometer. We emphasize that this population transfer is possible only through the coupling with the undiffracted state $\ket{1,0,0,0}$.

A key aspect of any matter-wave optical element is the transfer of a ``classical'' phase to the atoms \cite{Borde1989, Borde2001, CohenTannoudji1992}. In the case of light-pulse atom optics, this is the optical phase difference between excitation beams at the position of the atoms \cite{Kasevich1991}. To illustrate how these optical phases play a role for 2D atom optics, we consider the specific case of resonant fields ($\delta_x = \delta_y = 0$) and Rabi frequencies of identical magnitude ($\chi_{\mu} = |\chi| e^{i \phi_{\mu}}$). Here, $\phi_{\mu} = \bm{k}_\mu \cdot \bm{r} + \varphi_{\mu}$ is the total phase difference between Raman beams along the $\mu$-axis, with the atomic position denoted by $\bm{r}$ and the laser phase difference by $\varphi_{\mu} = \varphi_{1\mu} - \varphi_{2\mu}$. In a truncated basis with $\bm{\mathcal{C}}^{\rm T} = (\mathcal{C}_0,\mathcal{C}_x, \mathcal{C}_y)$, the 2D beamsplitter and mirror pulses can then be summarized by the following matrices
\begin{subequations}
\label{M2D}
\begin{align}
  & \mathbb{M}_{\rm 2D}(\tau) = -\begin{pmatrix}
    0 & \frac{i}{\sqrt{2}} e^{-i\phi_x} & \frac{i}{\sqrt{2}} e^{-i\phi_y} \\
    \frac{i}{\sqrt{2}} e^{i\phi_x} & -\frac{1}{2} & \frac{1}{2} e^{i(\phi_x - \phi_y)} \\
    \frac{i}{\sqrt{2}} e^{i\phi_y} & \frac{1}{2} e^{i(\phi_y - \phi_x)} & -\frac{1}{2}
  \end{pmatrix}, \\
  & \mathbb{M}_{\rm 2D}(2\tau) = -\begin{pmatrix}
    1 & 0 & 0 \\
    0 & 0 & e^{i(\phi_x - \phi_y)} \\
    0 & e^{i(\phi_y - \phi_x)} &  0
  \end{pmatrix}.
\end{align}
\end{subequations}
Here, the role of the optical phases becomes immediately clear. For atoms undergoing a two-photon transition from $\ket{1,0,0,0}$ to the diffracted state along axis $\mu$, the phase $\phi_{\mu}$ is imprinted on the wavepacket. This is a result of absorbing a photon from the field propagating along $\bm{\kappa}_{1\mu}$, followed by stimulated emission into the field along $\bm{\kappa}_{2\mu}$. Similarly, the phase $-\phi_{\mu}$ is imprinted when making the transition from the same diffracted state back to $\ket{1,0,0,0}$. Finally, atoms transferred between diffracted states acquire the phase $\pm(\phi_x - \phi_y)$. This arises because there is no direct coupling between $\ket{2,\hbar k_x,0,0}$ and $\ket{2,0,\hbar k_y,0}$---atoms must make a four-photon transition through the intermediate state $\ket{1,0,0,0}$ in a similar manner to double diffraction \cite{Leveque2009, Giese2013}.

\begin{figure}[!b]
  \centering
  \includegraphics[width=0.48\textwidth]{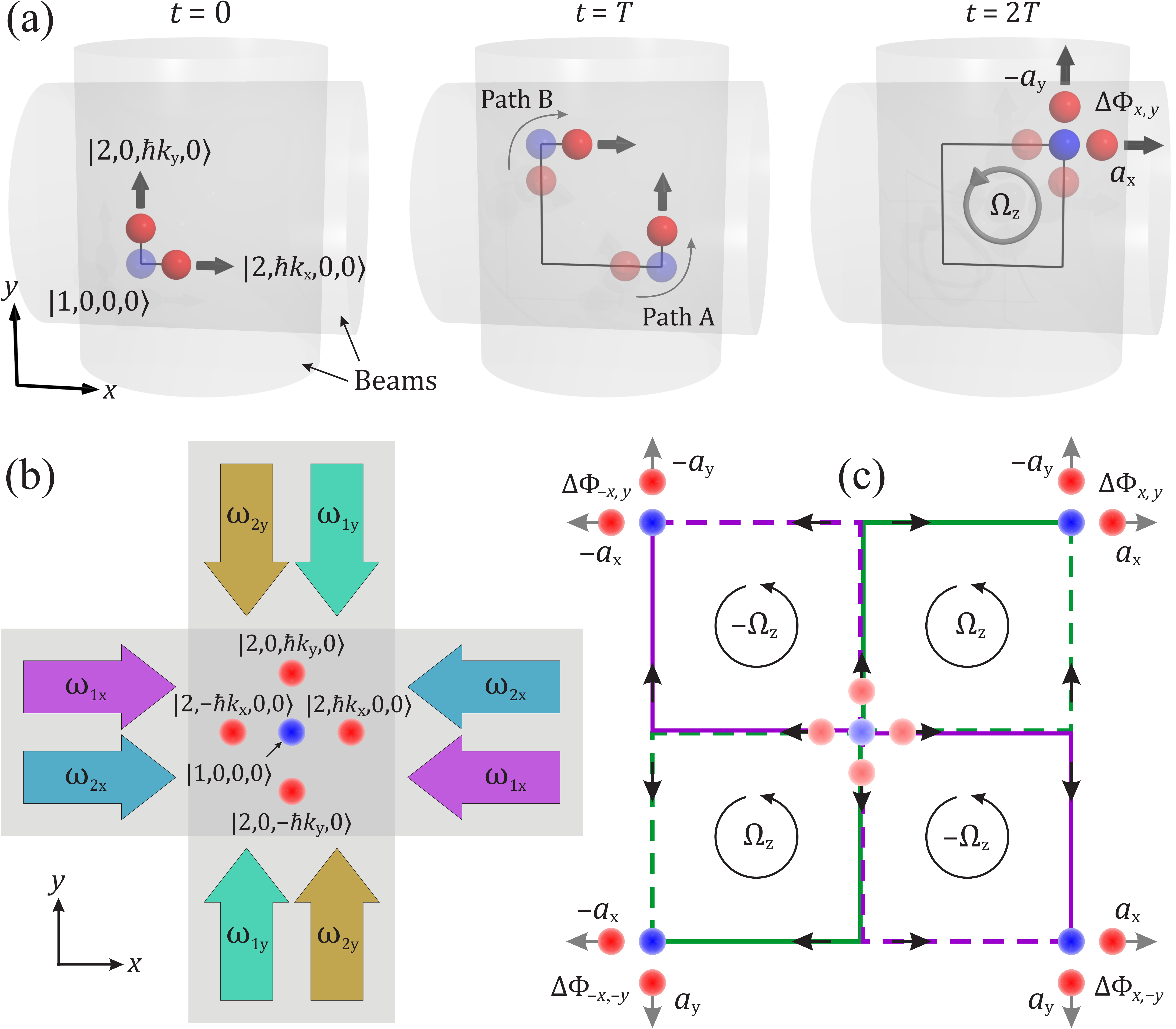}
  \caption{(a) A sequence of 2D atom-optical pulses constituting a 2D Mach-Zehnder interferometer. (b) A retro-reflected beam geometry enabling 2D double-diffraction atom optics, which symmetrically transfer $\pm\hbar k_{\mu}$ of momentum along each axis $\mu = x,y$. (c) Four simultaneous 2D Mach-Zehnder AIs derived from the same atomic source. Linear combinations of the four phase shifts allow one to isolate the three inertial components $a_x$, $a_y$ and $\Omega_z$ with increased sensitivity.}
  \label{fig:2D-MZ-Diagrams}
\end{figure}

A 2D Mach-Zehnder interferometer can be formed by combining a sequence of three 2D atom-optical pulses of duration $\tau - 2\tau - \tau$, each separated by an interrogation time $T$. Figure \ref{fig:2D-MZ-Diagrams}(a) shows the atomic trajectories associated with this new AI geometry, where atoms are split, reflected, and recombined along two spatial directions. A simple matrix representation of this process is obtained from the following product
\be
  \mathbb{M}_{\rm MZ} = \mathbb{M}_{\rm 2D}(\tau) \mathbb{U}_{\rm free}(T) \mathbb{M}_{\rm 2D}(2\tau) \mathbb{U}_{\rm free}(T) \mathbb{M}_{\rm 2D}(\tau),
\ee
where $\mathbb{U}_{\rm free}(T)$ is a unitary matrix describing the free evolution between laser interactions \cite{Cheinet2008}. For an atom initially in $\ket{1,0,0,0}$, and allowing for different optical phases $\phi_{\mu,i}$ during the $i^{\rm th}$ pulse, one can show that the two internal state populations---corresponding to the two complimentary output ports of the AI---are given by
\begin{align*}
  \label{2DFringes}
  |\!\bra{1} \mathbb{M}_{\rm MZ} \ket{1,0,0,0}\!|^2 & = \tfrac{1}{2} \left(1 \!+\! \cos\Delta\Phi\right) = |\mathcal{C}_0|^2, \\
  |\!\bra{2} \mathbb{M}_{\rm MZ} \ket{1,0,0,0}\!|^2 & = \tfrac{1}{2} \left(1 \!-\! \cos\Delta\Phi\right) = |\mathcal{C}_x|^2 \!+\! |\mathcal{C}_y|^2,
\end{align*}
where $\Delta\Phi = \Delta\phi_1 - 2\Delta\phi_2 + \Delta\phi_3$ is the total AI phase shift, with $\Delta\phi_i \equiv \phi_{x,i} - \phi_{y,i}$. We point out that the populations of the two diffracted states, $|\mathcal{C}_x|^2$ and $|\mathcal{C}_y|^2$, are identical and hence carry the same information.

The state-labelled architecture of this AI enables one to readout the two AI ports by spatial-integration using resonant fluorescence or absorption imaging \cite{Rocco2014}. Although the output ports are spatially separated, the 2D AI shown in \Fig \ref{fig:2D-MZ-Diagrams}(a) does not require a spatially-resolved detection system \cite{Dickerson2013, Sugarbaker2013, Hoth2016}. An interference fringe can be obtained from either port by scanning the optical phases---allowing one to probe for inertial effects.

The atomic trajectories shown in \Fig \ref{fig:2D-MZ-Diagrams}(a) give this 2D AI a unique sensitivity to inertial effects. Intuitively, since the two pathways enclose a rectangular spatial area in the $xy$-plane, the inertial phase is sensitive to the rotation component perpendicular to this plane, $\Omega_z$. This sensitivity is proportional to the area enclosed by the two pathways and does not require an initial velocity. In addition, when projected onto the $xt$- and $yt$-planes, these pathways enclose the same space-time area as a 1D Mach-Zehnder geometry---yielding sensitivity to the two acceleration components $a_x$ and $a_y$.

The full inertial dynamics resulting from the interference between any two atomic trajectories are encoded in the phase shift $\Delta\Phi$. We compute this phase shift for the 2D Mach-Zehnder geometry using the ABCD$\xi$ formalism developed by Bord\'{e} and Antoine \cite{Borde2001, Antoine2003a, Antoine2003b, Borde2004}. Briefly, $\Delta\Phi$ can be written as
\be
  \label{DeltaPhi}
  \Delta\Phi = \sum_{i=1}^N \Delta\bm{K}_i \cdot \bm{Q}_i + \Delta\varphi_i,
\ee
where $\Delta\bm{K}_i \equiv \bm{k}_{{\rm A},i} - \bm{k}_{{\rm B},i}$ is the difference between the effective wavevectors $\bm{k}_{{\rm A},i}$ and $\bm{k}_{{\rm B},i}$ associated with the momentum transfer from the $i^{\rm th}$ light pulse along paths ``A'' and ``B'', respectively. Similarly, $\bm{Q}_i \equiv \half\big(\bm{q}_{\rm A}(t_i) + \bm{q}_{\rm B}(t_i)\big)$ is the position on the mid-point trajectory, and $\Delta\varphi_i = \varphi_{{\rm A},i} - \varphi_{{\rm B},i}$ is a control parameter arising from the relative laser phases. The atomic position $\bm{q}$ and momentum $\bm{p}$ trajectories are computed from the solution to the classical equations of motion \cite{Antoine2003a}. Due to the symmetry of the Mach-Zehnder geometry (\ie $\bm{k}_{\rm A,2} + \bm{k}_{\rm B,2} = 0$), the phase shift is entirely determined by the choice of initial wavevectors $\bm{k}_{\rm A,1}$ and $\bm{k}_{\rm B,1}$ \footnote{The final wavevectors are predetermined by the initial wavevectors in order to close the two interferometer arms, \ie $\bm{k}_{\rm A,1} + \bm{k}_{\rm B,3} = 0$ and $\bm{k}_{\rm A,3} + \bm{k}_{\rm B,1} = 0$.}. In what follows, we label the AI phase shift with the subscript ``A,B'' which specifies both its geometry and initial wavevectors. To leading order in $T$, the generalized Mach-Zehnder phase shift is \cite{SupMat}
\be
  \label{DeltaPhiAB}
  \Delta\Phi_{\rm A,B} = \Delta\bm{K}_1 \!\cdot\! \left[\bm{a}
  + 2\left(\! \bm{v}_1 + \frac{\hbar}{m} \bm{K}_1 \!\right) \!\times\! \bm{\Omega} \right] T^2.
\ee
Here, $\bm{a} = (a_x,a_y,a_z)$ is the acceleration vector due to external motion and gravity, $\bm{\Omega} = (\Omega_x,\Omega_y,\Omega_z)$ is the rotation vector, $\bm{v}_1$ is the atomic velocity at the time of the first light pulse, $\bm{K}_1 \equiv \half \big(\bm{k}_{\rm A,1} + \bm{k}_{\rm B,1} \big)$ corresponds to the momentum transferred to the atom's center of mass by the first pulse, and we have omitted the control phases $\Delta\varphi_i$ for clarity. The first two terms in \Eq \eqref{DeltaPhiAB} correspond to the well-known first-order phase shift $\Delta\bm{K}_1\cdot(\bm{a} + 2\bm{v}_1\times\bm{\Omega})T^2$ which exhibits sensitivity to the components of $\bm{a}$ and the Coriolis acceleration $2\bm{v}_1\times\bm{\Omega}$ that are parallel to $\Delta\bm{K}_1$. The third term is a purely rotational phase which can be written as $2\frac{\hbar}{m} \big(\Delta\bm{K}_1 \times \bm{K}_1\big) \cdot \bm{\Omega} \, T^2$. We emphasize that this phase is not present in 1D light-pulse AIs where $\Delta\bm{K}_1 \times \bm{K}_1 = 0$. This key point leads to additional rotation and gravity gradient sensitivity \cite{SupMat} with multi-dimensional geometries that has not yet been exploited experimentally. In contrast to previous atomic gyroscopes \cite{Gustavson1997, Stockton2011, Barrett2014b, Tackmann2014, Rakholia2014, Dutta2016, Yao2018}, here an initial launch velocity is not required to achieve rotation sensitivity---instead this velocity is provided by the first 2D beamsplitter. This is advantageous for two reasons: (\emph{i}) the magnitude of this velocity kick can be as precise as the value of $k$ (typically better than one part in $10^9$), and (\emph{ii}) the direction of the kick can be changed by simply reversing the sign of $\bm{k}_{\rm A,1}$ and $\bm{k}_{\rm B,1}$. With a single atomic source, these features can then be exploited to suppress contributions from pure accelerations and initial velocities---which are the main sources of error in atomic gyroscopes \cite{Barrett2014b}.

\begin{table}[!b]
  \centering
  \small
  \begin{tabular}{cccccc}
    \hline
    \hline
      $\Delta\Phi_{x,y}$ & $\Delta\Phi_{-x,y}$ & $\Delta\Phi_{-x,-y}$ & $\Delta\Phi_{x,-y}$ & Sum phase \\
    \hline
      $+$ & $-$ & $-$ & $+$ & $4 k_x a_x^{\rm tot} T^2$ \\
      $-$ & $-$ & $+$ & $+$ & $4 k_y a_y^{\rm tot} T^2$ \\
      $+$ & $-$ & $+$ & $-$ & $8 \frac{\hbar}{m} k_x k_y \Omega_z T^2$ \\
    \hline
    \hline
  \end{tabular}
  \caption{Linear combinations of 2D Mach-Zehnder phases obtained from the four symmetric geometries shown in \Fig \ref{fig:2D-MZ-Diagrams}(c). Here, the laser phase contribution $\Delta\varphi_1 - 2\Delta\varphi_2 + \Delta\varphi_3$ cancels in the sum phase since it is common to all geometries.}
  \label{tab:DeltaPhiComb}
\end{table}

For the 2D Mach-Zehnder geometry shown in \Fig \ref{fig:2D-MZ-Diagrams}(a), with initial wavevectors in the $xy$-plane ($\bm{k}_{\rm A,1} = k_x \hat{\bm{x}}$, $\bm{k}_{\rm B,1} = k_y \hat{\bm{y}}$, $\Delta\bm{K}_1 \times \bm{K}_1 = k_x k_y \hat{\bm{z}}$), \Eq \eqref{DeltaPhiAB} gives
\be
  \label{DeltaPhixy}
  \Delta\Phi_{x,y} = k_x a_x^{\rm tot} T^2 - k_y a_y^{\rm tot} T^2 + \frac{2\hbar}{m} k_x k_y \Omega_z T^2,
\ee
where $\bm{a}^{\rm tot} \equiv \bm{a} + 2\bm{v}_1\times\bm{\Omega}$. Although this phase contains a mixture of different inertial effects, one can isolate each of them by using linear combinations of phases obtained from area-reversed geometries, as shown in Table \ref{tab:DeltaPhiComb}. Each phase $\Delta\Phi_{\rm A,B}$ can be obtained from a single measurement by employing double-diffraction \cite{Leveque2009, Malossi2010, Giese2013, Kuber2016, Ahlers2016} or double-single-diffraction \cite{Barrett2016a} atom optics in two dimensions \cite{SupMat}. Figures \ref{fig:2D-MZ-Diagrams}(b) and (c) display a scheme in which four simultaneous interferometers are generated from the same atomic source via 2D double diffraction pulses---enabling one to isolate $a_x$, $a_y$, and $\Omega_z$ in a single shot. Here, the phase readout requires spatial resolution of the adjacent interferometer ports, therefore the cloud diameter at the final beamsplitter must be less than the separation between adjacent clouds. This implies a sub-recoil-cooled source with an initial cloud size $\sigma_0 \ll 2 \hbar k_{\mu} T/m$ (\eg $\sigma_0 \ll 1$ mm for $T = 10$ ms). This scheme is well-suited to inertial navigation applications, where strong variations of rotations and accelerations between measurement cycles would compromise the common-mode rejection of a sequential measurement protocol. Additionally, with strongly-correlated measurements, one can reject both the phase noise between orthogonal Raman beams and common systematic effects.

\begin{figure}[!t]
  \centering
  \includegraphics[width=0.49\textwidth]{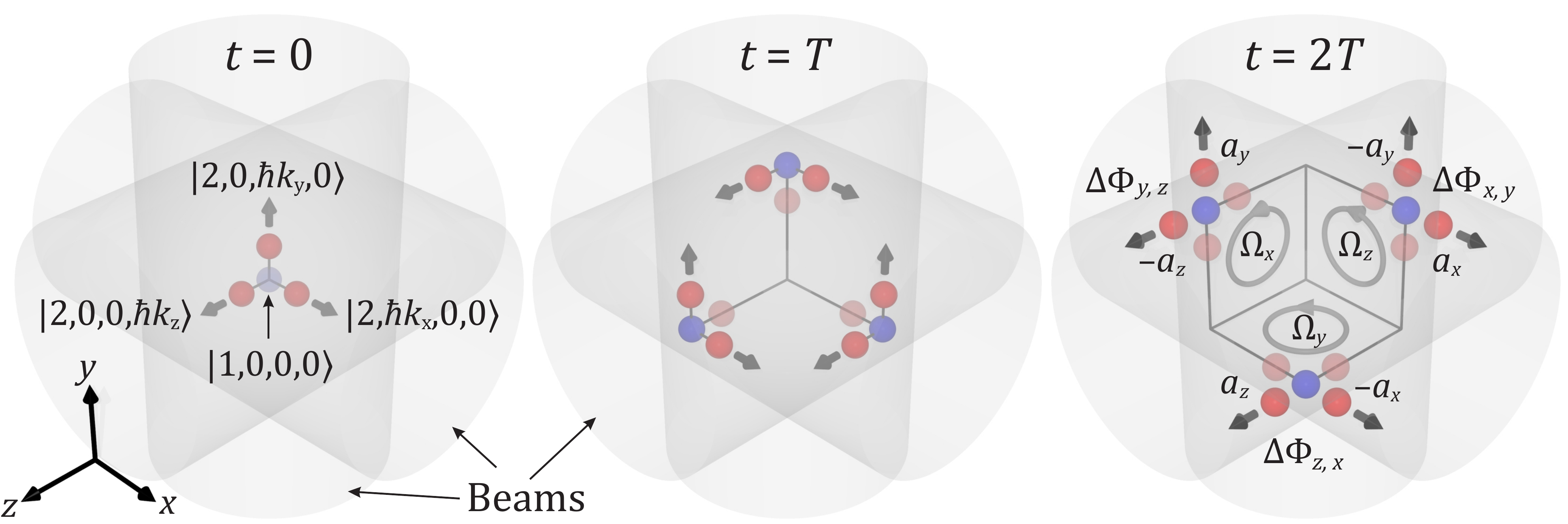}
  \caption{A sequence of 3D atom optical pulses generating three 2D Mach-Zehnder interferometers in mutually-orthogonal planes. Parasitic trajectories excited by the beams are not shown. Three sets of spatially-separated atomic clouds arrive at opposite corners of a cube where they can be read out simultaneously---yielding sensitivity to the full acceleration and rotation vectors.}
  \label{fig:3D-MZ-Diagrams}
\end{figure}

These principles can be extended to a 3D geometry, where three mutually-perpendicular pairs of Raman beams intersect to generate three separate 2D interferometers in orthogonal planes, as shown in \Fig \ref{fig:3D-MZ-Diagrams}. At $t = 0$, a 3D beamsplitter diffracts an atom initially in the undiffracted state $\ket{1,0,0,0}$ into three equal proportions traveling along $\hat{\bm{x}}$, $\hat{\bm{y}}$, and $\hat{\bm{z}}$. These three diffracted states continue along their respective axes ($\hat{\bm{\mu}}$) until $t = T$, when a 3D atom-optical mirror freezes the motion along $\hat{\bm{\mu}}$ and diffracts each wavepacket equally along the two directions orthogonal to $\hat{\bm{\mu}}$ \footnote{After the 3D mirror pulse, 1/9 of the population on each arm remains in the undiffracted state (not shown in \Fig \ref{fig:3D-MZ-Diagrams}), which leads to a loss of interference contrast.}. Finally, at $t = 2T$ the atoms intersect at three opposite corners of a cube, as shown in \Fig \ref{fig:3D-MZ-Diagrams}(c), where a recombination pulse transfers population from the diffracted states in each plane to an undiffracted one. Detection of the resulting 9 spatially-separated clouds yields sensitivity to the full acceleration and rotation vectors in a single shot. Individual inertial components can then be isolated in the same manner previously described for a 2D geometry---that is, by exciting symmetrically with double-diffraction pulses and imaging the clouds separately in each plane.

This 3D geometry allows one to easily construct a three-axis gyroscope. For instance, the combination of phases: $\Theta \equiv \Delta\Phi_{x,y} + \Delta\Phi_{y,z} + \Delta\Phi_{z,x}$, obtained from three corners of the cube, and $\Upsilon \equiv \Delta\Phi_{-x,y} + \Delta\Phi_{y,z} + \Delta\Phi_{z,-x}$, acquired by reversing the Raman wavevector on the $x$-axis, yields
\begin{subequations}
  \label{Phi3D}
\begin{align}
  \label{Phi3D-A}
  \Theta & = \frac{2\hbar}{m} (k_y k_z \Omega_x + k_x k_z \Omega_y + k_x k_y \Omega_z) T^2, \\
  \label{Phi3D-B}
  \Upsilon & = \frac{2\hbar}{m} (k_y k_z \Omega_x - k_x k_z \Omega_y - k_x k_y \Omega_z) T^2.
\end{align}
\end{subequations}
These phase combinations allow one to access $\Omega_x$ through the sum $\Theta + \Upsilon = 4\frac{\hbar}{m} k_y k_z \Omega_x T^2$. Hence, \Eqs \eqref{Phi3D} can be used as a building block to isolate each rotation component. A key point here is that both $\Theta$ and $\Upsilon$ arise from a single measurement of three simultaneous 2D interferometers in orthogonal planes. Yet they are each immune to spurious velocities, accelerations and laser phase noise, and hence can be combined to isolate a given rotation component. Since all quantities appearing in these rotation phases are precisely known, future gyroscopes based on this architecture could benefit from the same relative accuracy as cold-atom-based accelerometers \cite{Menoret2018}.


We have presented a novel approach for manipulating atomic wavepackets in multiple spatial dimensions. These new atom-optical tools can be utilized to generate simple 2D interferometers sensitive to inertial effects in 3D. More complex planar geometries involving 2D double diffraction pulses enable one to isolate two components of acceleration and one rotation in a single shot, while also rejecting laser phase noise and common systematic effects. Finally, we discussed an extension to a 3D geometry, where the full acceleration and rotation vectors can be retrieved. These concepts can easily be extended to other AI configurations involving four or more pulses \cite{Dubetsky2006, Cadoret2016}, which could be advantageous for applications such as multi-axis gravity gradiometry, gyroscopy, or gravitational wave detection. The sensitivity of these AIs could also benefit from multi-photon momentum transfer pulses \cite{Malinovsky2003, Clade2009, Kovachy2012, McDonald2013, Kotru2015, Jaffe2018}, which would aid the realization of a 3D inertial sensor in a compact volume. We anticipate that this work will influence future generations of quantum accelerometers and gyroscopes, and will offer new perspectives for inertial navigation systems.


This work is supported by the French national agencies ANR (l'Agence Nationale pour la Recherche), DGA (D\'{e}l\'egation G\'{e}n\'{e}rale de l'Armement) under the ANR-17-ASTR-0025-01 grant, IFRAF (Institut Francilien de Recherche sur les Atomes Froids), and action sp\'{e}cifique GRAM (Gravitation, Relativit\'{e}, Astronomie et M\'{e}trologie). P. Bouyer thanks Conseil R\'{e}gional d'Aquitaine for the Excellence Chair.


\bibliography{References}

\onecolumngrid

\setcounter{section}{0}
\setcounter{equation}{0}

\section*{Supplementary Material for\\``Multi-Dimensional Atom Optics and Interferometry''}


\section{3D atom optics based on Raman transitions}
\label{sec:3DAtomOptics}

Here, we describe the semi-classical formalism used to derive the dynamical equations for multi-dimensional beamsplitters and mirrors. Our approach closely follows previous work by Aspect \etal \cite{Aspect1989}, Moler \etal \cite{Moler1992}, and Giese \etal \cite{Giese2013}. We consider an atom in a superposition of states
\begin{align}
\begin{split}
  \ket{\psi(t)}
  & = \mathcal{A}_1(t) \ket{1,\bm{p}}
    + \mathcal{A}_{2x}(t) \ket{2,\bm{p} + \hbar\bm{k}_x}
    + \mathcal{A}_{2y}(t) \ket{2,\bm{p} + \hbar\bm{k}_y}
    + \mathcal{A}_{2z}(t) \ket{2,\bm{p} + \hbar\bm{k}_z} \\
  & + \mathcal{A}_{3x}(t) \ket{3,\bm{p} + \hbar\bm{\kappa}_{1x}}
    + \mathcal{A}_{3y}(t) \ket{3,\bm{p} + \hbar\bm{\kappa}_{1y}}
    + \mathcal{A}_{3z}(t) \ket{3,\bm{p} + \hbar\bm{\kappa}_{1z}},
\end{split}
\end{align}
which are labeled by their internal atomic energy, and the components of the atomic momentum vector $\bm{p}$ along orthogonal directions $\hat{\bm{\mu}} = \hat{\bm{x}}, \hat{\bm{y}}, \hat{\bm{z}}$. In this representation, the complex coefficients $\mathcal{A}_i(t)$ are time-varying state amplitudes that preserve normalization. The primary goal of this section is to derive expressions for these coefficients based on the Schr\"{o}dinger equation with an appropriate Hamiltonian $\mathcal{H}$
\be
  i\hbar \dot{\bm{\mathcal{A}}} = \mathcal{H} \bm{\mathcal{A}}(t),
\ee
where the vector $\bm{\mathcal{A}}^{\rm T} = (\mathcal{A}_1, \mathcal{A}_{2x}, \mathcal{A}_{2y}, \mathcal{A}_{2z}, \mathcal{A}_{3x}, \mathcal{A}_{3y}, \mathcal{A}_{3z})$ represents the superposition state at time $t$. The ground states $\ket{1}$ and $\ket{2}$ are coupled through an intermediate excited state $\ket{3}$ by three pairs of monochromatic laser fields. We label each field parameter with the pair of subscripts $n\mu$, where $n = 1, 2$ corresponds to the internal ground state to which the field frequency is near resonant (\ie the transition $\ket{n} \to \ket{3}$), and $\mu = x,y,z$ indicates the axis along which the laser field is traveling. The total electric field can then be written as the following sum of travelling waves
\begin{align}
\begin{split}
  \bm{E}(\bm{r},t) & = \frac{1}{2} \sum_{n,\mu} \bm{E}_{n\mu} e^{i(\bm{\kappa}_{n\mu} \cdot \bm{r} - \omega_{n\mu} t + \varphi_{n\mu})} + \mbox{c.c.},
\end{split}
\end{align}
with counter-propagating, orthogonal wavevectors $\bm{\kappa}_{n\mu} = (-1)^{n+1} \kappa_{n\mu} \hat{\bm{\mu}}$, frequencies $\omega_{n\mu}$ and phases $\varphi_{n\mu}$. The effective wavevectors $\bm{k}_{\mu} = \bm{\kappa}_{1\mu} - \bm{\kappa}_{2\mu}$ appearing in the basis states describe the total momentum transferred from the laser fields to the atom along each direction.

We adopt the usual semi-classical Hamiltonian with contributions from the bare atom and the atom-field interaction \cite{Moler1992}
\be
  \mathcal{H} = \mathcal{H}_{\rm A} + V, \;\;\;\;\;\;
  \mathcal{H}_{\rm A} = \frac{\bm{P}^2}{2m} + \hbar \sum_{j = 1}^3 \omega^{\rm A}_j \ket{j}\bra{j}, \;\;\;\;\;\;
  V = -\bm{D} \cdot \bm{E}(\bm{r},t),
\ee
where $\bm{P}$ is the momentum operator, $\bm{D}$ is the electric dipole operator, and $\hbar \omega^{\rm A}_j$ is the internal energy of state $\ket{j}$. We ignore spontaneous emission by assuming the detuning from the optical transition is much larger than the natural linewidth of the excited state. Additionally, we assume that fields $\bm{E}_{1\mu}$ couple only $\ket{1}$ and $\ket{3}$, and fields $\bm{E}_{2\mu}$ couple only $\ket{2}$ and $\ket{3}$, which is a good approximation when the hyperfine splitting is large compared to the optical detuning. With this simplification, the atom-field interaction can be written as
\begin{align}
\begin{split}
  V
  & = \hbar\chi_{1x} e^{-i\omega_{1x} t} \ket{3,\bm{p}+\hbar\bm{\kappa}_{1x}} \bra{1,\bm{p}}
  + \hbar\chi_{2x} e^{-i\omega_{2x} t} \ket{3,\bm{p}+\hbar\bm{\kappa}_{1x}}
  \bra{2,\bm{p}+\hbar\bm{k}_x} \\
  & + \hbar\chi_{1y} e^{-i\omega_{1y} t} \ket{3,\bm{p}+\hbar\bm{\kappa}_{1y}} \bra{1,\bm{p}}
  + \hbar\chi_{2y} e^{-i\omega_{2y} t} \ket{3,\bm{p}+\hbar\bm{\kappa}_{1y}}
  \bra{2,\bm{p}+\hbar\bm{k}_y} \\
  & + \hbar\chi_{1z} e^{-i\omega_{1z} t} \ket{3,\bm{p}+\hbar\bm{\kappa}_{1z}} \bra{1,\bm{p}}
  + \hbar\chi_{2z} e^{-i\omega_{2z} t} \ket{3,\bm{p}+\hbar\bm{\kappa}_{1z}}
  \bra{2,\bm{p}+\hbar\bm{k}_z} + \mbox{c.c.}
\end{split}
\end{align}
Here, the one-photon Rabi frequencies for each field are given by $\chi_{n\mu} = -\frac{1}{2\hbar} \bra{3}\bm{D}\cdot\bm{E}_{n\mu}\ket{n} e^{i\varphi_{n\mu}}$, and we have used the identity
\be
  e^{\pm i\bm{k}\cdot\bm{r}} = \int \ket{\bm{p} \pm \hbar \bm{k}}\bra{\bm{p}} \dd \bm{p}.
\ee
We have omitted contributions from parasitic states, such as $\ket{1,\bm{p} \pm \hbar(\bm{\kappa}_{1x} - \bm{\kappa}_{1y})}$ and $\ket{2,\bm{p} + \hbar(\bm{\kappa}_{1x} + \bm{\kappa}_{2y})}$, because with an appropriate choice of frequencies (\ie $|\omega_{n\mu} - \omega_{n\nu}| \gg |\chi_{n\mu}|, |\chi_{n\nu}|$ for $\mu \neq \nu$), these states can be made non-resonant with the applied fields. For the moment we also omit states that require multiple two-photon processes, such as $\ket{1,\bm{p} \pm \hbar (\bm{k}_x + \bm{k}_y)}$ and $\ket{1,\bm{p} \pm 2\hbar \bm{k}_x}$, but we revisit these processes in \Sec \ref{sec:DoubleDiffraction2DAtomOptics}.

Next, we transform into the interaction representation, where the state amplitudes rotate at a frequency corresponding to their internal energy, by performing the following unitary operation
\begin{subequations}
\begin{align}
  \bm{\mathcal{B}}
  & = e^{i \mathcal{H}_{\rm A} t/\hbar} \bm{\mathcal{A}}, \\
  \label{Bdot}
  \dot{\bm{\mathcal{B}}}
  & = \frac{i}{\hbar} \mathcal{H}_{\rm A} e^{i \mathcal{H}_{\rm A} t/\hbar} \bm{\mathcal{A}} + e^{i \mathcal{H}_{\rm A} t/\hbar} \dot{\bm{\mathcal{A}}} = -\frac{i}{\hbar} \mathcal{H}_{\rm int} \bm{\mathcal{B}}.
\end{align}
\end{subequations}
In doing so, we obtain a simpler set of equations for the coefficients $\mathcal{B}_i(t)$, which contain no diagonal elements and are governed by the Hamiltonian in the interaction representation
\be
  \mathcal{H}_{\rm int} = e^{i \mathcal{H}_{\rm A} t/\hbar} V e^{-i \mathcal{H}_{\rm A} t/\hbar}.
\ee
After some algebra, one can show that \Eq \eqref{Bdot} reduces to the following system
\begin{subequations}
\begin{align}
  \label{dotB1}
  \dot{\mathcal{B}}_{1}
  & = -i(\chi_{1x}^* e^{i\Delta_x t} \mathcal{B}_{3x} + \chi_{1y}^* e^{i\Delta_y t} \mathcal{B}_{3y} + \chi_{1z}^* e^{i\Delta_z t} \mathcal{B}_{3z}), \\
  \label{dotB2x}
  \dot{\mathcal{B}}_{2x}
  & = -i\chi_{2x}^* e^{i(\Delta_x - \delta_x)t} \mathcal{B}_{3x}, \\
  \label{dotB2y}
  \dot{\mathcal{B}}_{2y}
  & = -i\chi_{2y}^* e^{i(\Delta_y - \delta_y)t} \mathcal{B}_{3y}, \\
  \label{dotB2z}
  \dot{\mathcal{B}}_{2z}
  & = -i\chi_{2z}^* e^{i(\Delta_z - \delta_z)t} \mathcal{B}_{3z}, \\
  \label{dotB3x}
  \dot{\mathcal{B}}_{3x}
  & = -i(\chi_{1x} e^{-i\Delta_x t} \mathcal{B}_{1} + \chi_{2x} e^{-i(\Delta_x - \delta_x)t} \mathcal{B}_{2x}), \\
  \label{dotB3y}
  \dot{\mathcal{B}}_{3y}
  & = -i(\chi_{1y} e^{-i\Delta_y t} \mathcal{B}_{1} + \chi_{2y} e^{-i(\Delta_y - \delta_y)t} \mathcal{B}_{2y}), \\
  \label{dotB3z}
  \dot{\mathcal{B}}_{3z}
  & = -i(\chi_{1z} e^{-i\Delta_z t} \mathcal{B}_{1} + \chi_{2z} e^{-i(\Delta_z - \delta_z)t} \mathcal{B}_{2z}).
\end{align}
\end{subequations}
Here, $\Delta_{\mu}$ and $\delta_{\mu}$ are the detunings for fields propagating along axis $\mu$ corresponding to one- and two-photon processes, respectively. They are defined as
\begin{subequations}
\begin{align}
  \label{Delta_mu}
  \Delta_{\mu} & = \omega_{1\mu} - \omega^{\rm A}_{31} - \omega_{1\mu}^{\rm D} - \omega_{1\mu}^{\rm R}, \\
  \label{delta_mu}
  \delta_{\mu} & = \delta_{\mu}^{\rm L} - \omega^{\rm A}_{21} - \delta_{\mu}^{\rm D} - \delta_{\mu}^{\rm R},
\end{align}
\end{subequations}
where $\omega^{\rm A}_{ij} = \omega^{\rm A}_i - \omega^{\rm A}_j$ is the atomic transition frequency between $\ket{i}$ and $\ket{j}$, $\omega_{n\mu}^{\rm D} = \bm{\kappa}_{n\mu} \!\cdot\! \bm{v}$ is a Doppler shift, and $\omega_{n\mu}^{\rm R} = \hbar \kappa_{n\mu}^2/2m$ is a recoil frequency. The two-photon detuning \eqref{delta_mu} contains similar terms for the laser frequency difference, Doppler effect and recoil shift
\be
  \!\!
  \delta_{\mu}^{\rm L} = \omega_{1\mu} - \omega_{2\mu}, \;\;\;
  \delta_{\mu}^{\rm D} = \bm{k}_{\mu} \!\cdot\! \bm{v}, \;\;\;
  \delta_{\mu}^{\rm R} = \frac{\hbar k_{\mu}^2}{2m}.
\ee

In deriving the equations for $\bm{\mathcal{B}}_i(t)$, we ignored effects due to spontaneous emission by assuming $|\Delta_{\mu}|$ is much larger than the atomic linewidth. Additionally, we assumed that $|\Delta_{\mu}| \gg |\chi_{n\mu}|, |\delta_{\mu}|$ such that the ground states in \Eqs \eqref{dotB3x}--\eqref{dotB3z} evolve much more slowly than the excited states, which can then be eliminated adiabatically by integrating their equations directly
\begin{subequations}
\begin{align}
  \label{B3x}
  \mathcal{B}_{3x}(t)
  & \approx \frac{e^{-i\Delta_x t}}{\Delta_x} \left[ \chi_{1x} \mathcal{B}_{1} + \chi_{2x} e^{i\delta_x t} \mathcal{B}_{2x} \right], \\
  \label{B3y}
  \mathcal{B}_{3y}(t)
  & \approx \frac{e^{-i\Delta_y t}}{\Delta_y} \left[ \chi_{1y} \mathcal{B}_{1} + \chi_{2y} e^{i\delta_y t} \mathcal{B}_{2y} \right], \\
  \label{B3z}
  \mathcal{B}_{3z}(t)
  & \approx \frac{e^{-i\Delta_z t}}{\Delta_z} \left[ \chi_{1z} \mathcal{B}_{1} + \chi_{2z} e^{i\delta_z t} \mathcal{B}_{2z} \right].
\end{align}
\end{subequations}
By reinserting these expressions into \Eqs \eqref{dotB1}--\eqref{dotB2z}, we obtain an effective four-level system for $\bm{\mathcal{B}}^{\rm T} = (\mathcal{B}_1, \mathcal{B}_{2x}, \mathcal{B}_{2y}, \mathcal{B}_{2z})$
\be
  \label{dotB}
  \dot{\bm{\mathcal{B}}} = -i \begin{pmatrix}
    \chi_{1}^{\rm AC} & \chi_x^* e^{i\delta_x t} & \chi_y^* e^{i\delta_y t} & \chi_z^* e^{i\delta_z t} \\
    \chi_x e^{-i\delta_x t} & \chi_{2x}^{\rm AC} & 0 & 0 \\
    \chi_y e^{-i\delta_y t} & 0 & \chi_{2y}^{\rm AC} & 0 \\
    \chi_z e^{-i\delta_z t} & 0 & 0 & \chi_{2z}^{\rm AC} \\
  \end{pmatrix} \bm{\mathcal{B}}
\ee
where $\chi_{\mu} = \frac{\chi_{1\mu}\chi_{2\mu}^*}{\Delta_\mu}$ is a two-photon Rabi frequency, $\chi_{n\mu}^{\rm AC} = \frac{|\chi_{n\mu}|^2}{\Delta_\mu}$ is a light shift, and $\chi_1^{\rm AC} \equiv \chi_{1x}^{\rm AC} + \chi_{1y}^{\rm AC} + \chi_{1z}^{\rm AC}$. We now make a unitary transformation to obtain a matrix containing only constant coefficients. To achieve this, we introduce a new vector of coefficients $\bm{\mathcal{C}}$ as defined by
\be
  \label{Cdef}
  \bm{\mathcal{C}} = \begin{pmatrix}
    1 & 0 & 0 & 0 \\
    0 & e^{i \delta_x t} & 0 & 0 \\
    0 & 0 & e^{i \delta_y t} & 0 \\
    0 & 0 & 0 & e^{i \delta_z t} \\
   \end{pmatrix} e^{i \chi_1^{\rm AC} t} \bm{\mathcal{B}},
\ee
which is obtained by combining two separate unitary transformations: one to remove the diagonal light-shift elements in \Eq \eqref{dotB}, and one to remove the $e^{i\delta_{\mu} t}$ factors in the off-diagonal elements. Combining \Eqs \eqref{dotB} and \eqref{Cdef}, one can show that
\be
  \dot{\bm{\mathcal{C}}} = -i \begin{pmatrix}
    0 & \chi_x^* & \chi_y^* & \chi_z^* \\
    \chi_x & -\tilde{\delta}_x & 0 & 0 \\
    \chi_y & 0 & -\tilde{\delta}_y & 0 \\
    \chi_z & 0 & 0 & -\tilde{\delta}_z
   \end{pmatrix} \bm{\mathcal{C}},
\ee
where $\tilde{\delta}_{\mu} \equiv \delta_{\mu} - \delta_{\mu}^{\rm AC}$, and $\delta_{\mu}^{\rm AC} = \chi_{2\mu}^{\rm AC} - \chi_1^{\rm AC}$ is a differential light shift. This simple form can be solved by direct integration
\be
  \label{C(t)2}
  \bm{\mathcal{C}}(t) = \underbrace{\exp\left[ -i \begin{pmatrix}
    0 & \chi_x^* & \chi_y^* & \chi_z^* \\
    \chi_x & -\tilde{\delta}_x & 0 & 0 \\
    \chi_y & 0 & -\tilde{\delta}_y & 0 \\
    \chi_z & 0 & 0 & -\tilde{\delta}_z
   \end{pmatrix} t \right]}_{\mathbb{M}_{\rm 3D}(t)} \bm{\mathcal{C}}(0).
\ee
This solution does not simplify to a more convenient analytical form, except for the specific case of identical detunings ($\tilde{\delta}_x = \tilde{\delta}_y = \tilde{\delta}_z \equiv \tilde{\delta}$). Then the matrix exponential $\mathbb{M}_{\rm 3D}(t)$ in \Eq \eqref{C(t)2} reduces to
\begin{align}
\begin{split}
  \label{MatrixExpExact}
  \mathbb{M}_{\rm 3D}(t) & = \left[ \begin{pmatrix}
     \chi_{xyz}^2 & 0 & 0 & 0 \\[0.1cm]
     0 & |\chi_x|^2 & \chi_x^* \chi_y & \chi_x^* \chi_z \\[0.2cm]
     0 & \chi_y^* \chi_x & |\chi_y|^2 & \chi_y^* \chi_z \\[0.2cm]
     0 & \chi_z^* \chi_x & \chi_z^* \chi_y & |\chi_z|^2 \\[0.2cm]
  \end{pmatrix} \frac{\Theta(t)}{\chi_{xyz}^2} +
  i\begin{pmatrix}
    -\tilde{\delta} & -2\chi_x & -2\chi_y & -2\chi_z \\
    -2\chi_x^* & \frac{|\chi_x|^2}{\chi_{xyz}^2} \tilde{\delta} & \frac{\chi_x^* \chi_y}{\chi_{xyz}^2} \tilde{\delta} & \frac{\chi_x^* \chi_z}{\chi_{xyz}^2} \tilde{\delta} \\
    -2\chi_y^* & \frac{\chi_y^* \chi_x}{\chi_{xyz}^2} \tilde{\delta} & \frac{|\chi_y|^2}{\chi_{xyz}^2} \tilde{\delta} & \frac{\chi_y^* \chi_z}{\chi_{xyz}^2} \tilde{\delta} \\
    -2\chi_z^* & \frac{\chi_z^* \chi_x}{\chi_{xyz}^2} \tilde{\delta} & \frac{\chi_z^* \chi_y}{\chi_{xyz}^2} \tilde{\delta} & \frac{|\chi_z|^2}{\chi_{xyz}^2} \tilde{\delta} \\
  \end{pmatrix} \frac{\Lambda(t)}{2\Omega_{\rm Rabi}}
  \right] e^{i \tilde{\delta} t/2} \\
  & + \begin{pmatrix}
    0 & 0 & 0 & 0 \\[0.1cm]
    0 & |\chi_y|^2 + |\chi_z|^2 & \chi_x^* \chi_y & \chi_x^* \chi_z \\[0.1cm]
    0 & \chi_y^* \chi_x & |\chi_x|^2 + |\chi_z|^2 & \chi_y^* \chi_z \\[0.1cm]
    0 & \chi_z^* \chi_x & \chi_z^* \chi_y & |\chi_x|^2 + |\chi_y|^2 \\[0.1cm]
  \end{pmatrix} \frac{e^{i \tilde{\delta} t}}{\chi_{xyz}^2},
\end{split}
\end{align}
where $\Theta(t) \equiv \cos(\Omega_{\rm Rabi} t)$ and $\Lambda(t) \equiv \sin(\Omega_{\rm Rabi} t)$, and the effective Rabi frequency---the rate at which the population in the undiffracted state oscillates as a function of the interaction time---is given by
\be
  \Omega_{\rm Rabi} = \frac{1}{2} \sqrt{\tilde{\delta}^2 + 4 \chi_{xyz}^2}, \;\;\;\;
  \chi_{xyz} = \sqrt{|\chi_x|^2 + |\chi_y|^2 + |\chi_z|^2}.
\ee
For on-resonance beams ($\tilde{\delta} = 0$) and Rabi frequencies with identical magnitudes ($\chi_{\mu} = |\chi|e^{i\phi_\mu}$, where $\phi_\mu = \varphi_{1\mu} - \varphi_{2\mu}$), the effective Rabi frequency reduces to $\Omega_{\rm Rabi} = \sqrt{3}|\chi|$, and $\mathbb{M}_{\rm 3D}(t)$ becomes
\be
  \label{MatrixExpExact2}
  \mathbb{M}_{\rm 3D}(t) = \begin{pmatrix}
    \Theta(t) & -\frac{i}{\sqrt{3}} \Lambda(t) e^{-i\phi_x} &
    -\frac{i}{\sqrt{3}} \Lambda(t) e^{-i\phi_y} &
    -\frac{i}{\sqrt{3}} \Lambda(t) e^{-i\phi_z} \\[0.1cm]
    -\frac{i}{\sqrt{3}} \Lambda(t) e^{i\phi_x} &
    \frac{1}{3}(\Theta(t) + 2) &
    \frac{1}{3}(\Theta(t) - 1) e^{i(\phi_x-\phi_y)} &
    \frac{1}{3}(\Theta(t) - 1) e^{i(\phi_x-\phi_z)} \\[0.1cm]
    -\frac{i}{\sqrt{3}} \Lambda(t) e^{i\phi_y} &
    \frac{1}{3}(\Theta(t) - 1) e^{i(\phi_y-\phi_x)} &
    \frac{1}{3}(\Theta(t) + 2) &
    \frac{1}{3}(\Theta(t) - 1) e^{i(\phi_y-\phi_z)} \\[0.1cm]
    -\frac{i}{\sqrt{3}} \Lambda(t) e^{i\phi_z} &
    \frac{1}{3}(\Theta(t) - 1) e^{i(\phi_z - \phi_x)} &
    \frac{1}{3}(\Theta(t) - 1) e^{i(\phi_z - \phi_y)} &
    \frac{1}{3}(\Theta(t) + 2) \\[0.1cm]
  \end{pmatrix}.
\ee

\section{2D Double-diffraction atom optics}
\label{sec:DoubleDiffraction2DAtomOptics}

In this Section, we discuss the principles of double-diffraction atom optics in two dimensions. As the name suggests, double diffraction refers to an excitation analogous to two simultaneous single-diffraction events, which produces a symmetric $2\hbar \bm{k}$ momentum-space splitting, where $\bm{k} \simeq 2\bm{\kappa}$ is the effective two-photon wavevector. In the 1D double-diffraction scheme, an atom initially in momentum state $\ket{\bm{p}}$ is coupled to the two symmetric states $\ket{\bm{p}+\hbar\bm{k}}$ and $\ket{\bm{p}-\hbar\bm{k}}$ via two pairs of counter-propagating laser fields \cite{Leveque2009, Giese2013}. Although double diffraction has been demonstrated with both Raman \cite{Leveque2009, Malossi2010} and Bragg \cite{Kuber2016, Ahlers2016} pulses in one dimension, to the best of our knowledge its generalization to multiple dimensions has not yet been considered.

\begin{figure}[!t]
  \centering
  \includegraphics[width=0.75\textwidth]{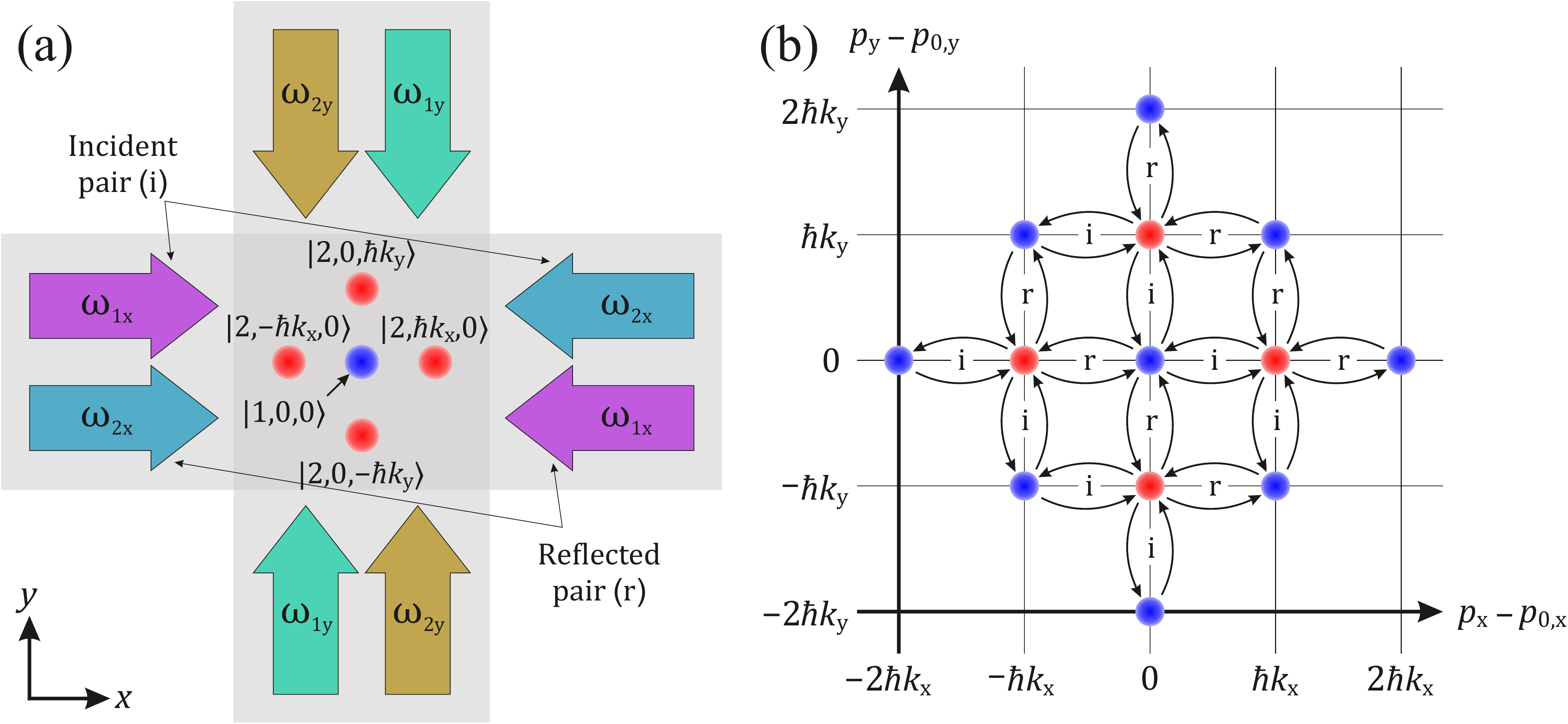}
  \caption{(a) Beam geometry for 2D double Raman diffraction consisting of two pairs of counter-propagating beams along each orthogonal direction. The optical frequencies are the same as in \Fig 1 in the main text. The target states for atom optics pulses are displayed in the region of overlap between all four beams. (b) Momentum-space representation of the full set of states considered in 2D double-diffraction simulations. Blue and red circles indicate internal states $\ket{1}$ and $\ket{2}$, respectively, and arrows signify the coupling between states considered in the model. Arrow labels ``i'' and ``r'' correspond, respectively, to couplings induced by the incident and reflected field pairs shown in (a).}
  \label{fig:2DAtomOptics-DD-Geometry}
\end{figure}

To achieve symmetric momentum splitting in 2D, we consider a geometry of retro-reflected beams in the $xy$-plane as shown in \Fig \ref{fig:2DAtomOptics-DD-Geometry}(a). Taking the $x$-axis as an example, the two pairs of counter-propagating beams with frequencies $\omega_{1x}$ and $\omega_{2x}$ (associated with effective wavevectors $\pm \bm{k}_x$) drive transitions between the undiffracted state $\ket{1,\bm{p}}$ and the horizontally diffracted states $\ket{2,\bm{p} \pm \hbar \bm{k}_x}$. To achieve this, the two pairs of beams must be simultaneously resonant with the two diffracted states. This is possible only for atoms with near-zero velocity along the $x$-axis, that is, velocities for which $v_x \ll \hbar k_x/m$. A similar process occurs along the $y$-axis, coupling the states $\ket{1,\bm{p}}$ and $\ket{2,\bm{p}\pm\hbar \bm{k}_y}$ for velocities $v_y \ll \hbar k_y/m$.

The set of laser frequencies that drive transitions $\ket{1,\bm{p}} \leftrightarrow \ket{2,\bm{p}\pm\hbar \bm{k}_x}$ are also near resonant with $\ket{2,\bm{p}+\hbar \bm{k}_y} \leftrightarrow \ket{1,\bm{p}+\hbar(\bm{k}_y\pm\bm{k}_x)}$ and $\ket{2,\bm{p}-\hbar \bm{k}_y} \leftrightarrow \ket{1,\bm{p} - \hbar(\bm{k}_y \pm \bm{k}_x)}$ [see \Fig \ref{fig:2DAtomOptics-DD-Geometry}(b)]. This is because the target states ($\ket{2,\bm{p}\pm\hbar \bm{k}_x}$ and $\ket{2,\bm{p}\pm\hbar \bm{k}_y}$) differ in kinetic energy from the parasitic states $\ket{1,\bm{p} \pm \hbar(\bm{k}_x \pm \bm{k}_y)}$ by only a Doppler shift ($\pm \delta_{\mu}^{\rm D} = \pm \bm{k}_{\mu} \cdot \bm{v}$) and the two-photon recoil frequency ($\delta_{\mu}^{\rm R} = \hbar k_{\mu}^2/2m$). However, for atoms involved in the diffraction process $|\delta_{\mu}^{\rm D}| \ll \delta_{\mu}^{\rm R}$ due to the condition imposed on the velocities---making the frequency separation between these states relatively small ($\delta_{\mu}^{\rm R} \simeq 2\pi \times 15$ kHz for $^{87}$Rb atoms). Atoms can accumulate in parasitic states from multiple two-photon processes and, although they do not participate in the 2D interferometers shown in \Fig 3(c) in the main text, these states can play an important role in the system dynamics.

To account for these higher-order processes, we consider a basis consisting of states $\ket{1,n_x\hbar k_x,n_x\hbar k_y}$ and $\ket{2,m_x\hbar k_x,m_y\hbar k_y}$ with integers $n_{\mu}, m_{\mu}$ for which $|n_x| + |n_y| = 0$ or $2$, and $|m_x| + |m_y| = 1$. These sums correspond to the total number of two-photon kicks allowed by the model for each internal state. The atomic wavepacket can then be written as a superposition of these states
\be
  \ket{\psi(t)}
  = \sum_{n_x,n_y} \mathcal{A}_{1,n_x,n_y}(t) \ket{1,n_x\hbar k_x,n_y\hbar k_y} + \!\! \sum_{m_x,m_y} \mathcal{A}_{2,m_x,m_y}(t) \ket{2,m_x\hbar k_x,m_y\hbar k_y},
\ee
where $\mathcal{A}_{j,m,n}(t)$ is a time-dependent state amplitude analogous to those presented in \Sec \ref{sec:3DAtomOptics}. The momentum-space representation of this basis is shown in \Fig \ref{fig:2DAtomOptics-DD-Geometry}(b). One can then derive the dynamical equations for the coupled state amplitudes following the same procedure outlined in \Sec \ref{sec:3DAtomOptics}.

\begin{figure}[!t]
  \centering
  \includegraphics[width=0.85\textwidth]{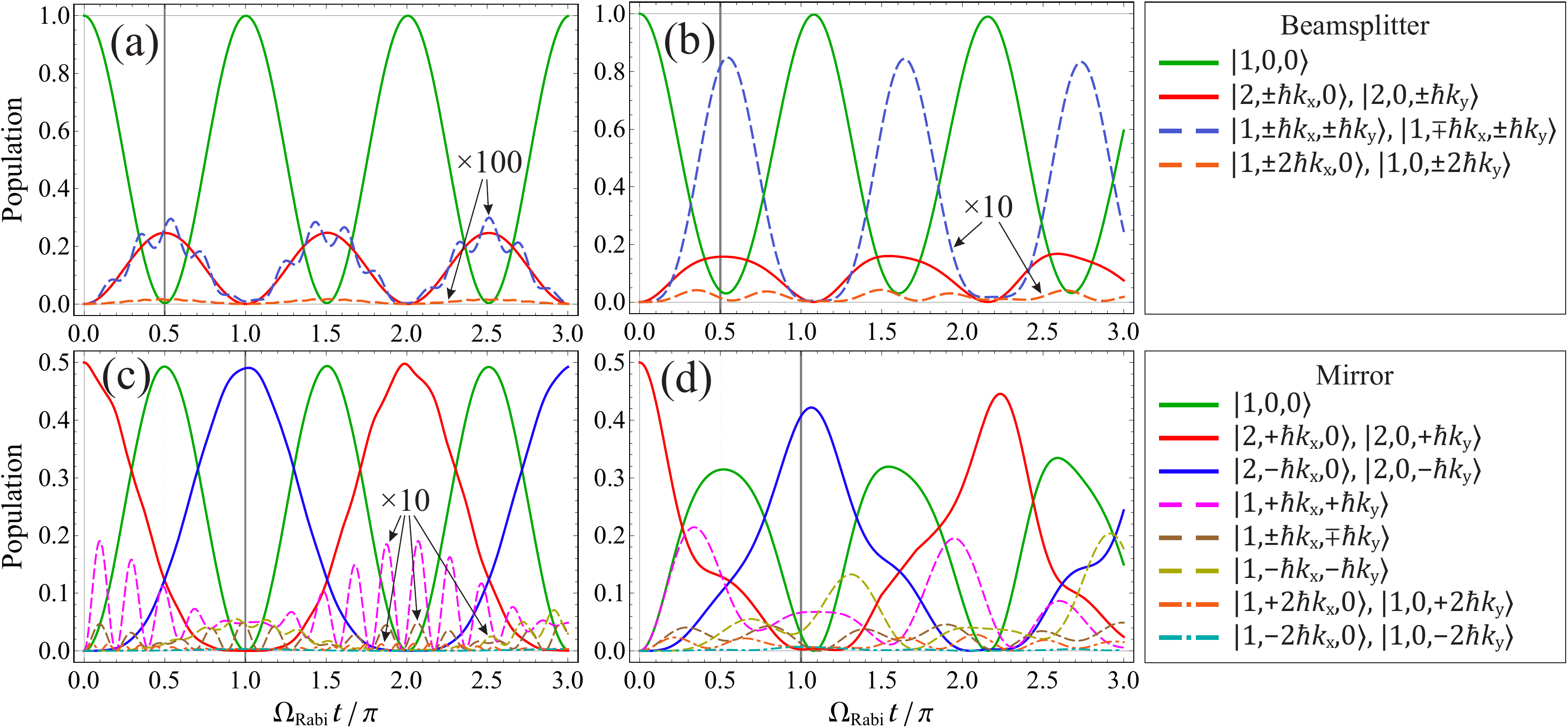}
  \caption{Rabi oscillations between states for 2D double Raman diffraction. Here, we consider the simple case of resonant beams with equal intensities incident on atoms with zero initial velocity. Light shifts are ignored for simplicity. (a,b) 2D double-diffraction beamsplitter for atoms initially in $\ket{1,0,0}$, where the target state is an equal superposition of $\ket{2,\pm\hbar k_x,0}$ and $\ket{2,0,\pm\hbar k_y}$. (c,d) 2D double-diffraction mirror for atoms initially in $\frac{1}{\sqrt{2}} \ket{2,\hbar k_x,0} + \frac{1}{\sqrt{2}} \ket{2,0,\hbar k_y}$. Here, the target state is $\frac{1}{\sqrt{2}} \ket{2,-\hbar k_x,0} + \frac{1}{\sqrt{2}} \ket{2,0,-\hbar k_y}$. Different beam intensities are used for each column: (a,c) $\Omega_{\rm Rabi} = 0.1\,\delta_{\mu}^{\rm R}$; (b,d) $\Omega_{\rm Rabi} = 0.5\,\delta_{\mu}^{\rm R}$. For clarity, some curves have been scaled up by factors of 10 or 100 as indicated in the figures.}
  \label{fig:2DAtomOptics-DD-RabiOsc}
\end{figure}

To illustrate the efficiency of 2D double-diffraction atom optics, we have numerically solved these dynamical equations under various conditions. Figure \ref{fig:2DAtomOptics-DD-RabiOsc} shows the corresponding Rabi oscillations in the population of each state as a function of the interaction time with the lasers. Here, we focus on the ideal case of beams with equal intensities, and atoms with zero initial velocity such that all beams are simultaneously resonant. When the effective Rabi frequency $\Omega_{\rm Rabi}$ is small compared to the recoil frequency, we find near-perfect efficiency for 2D double-diffraction atom optics, as shown in \Figs \ref{fig:2DAtomOptics-DD-RabiOsc}(a,c). With $\Omega_{\rm Rabi} = 0.1\,\delta_{\mu}^{\rm R}$, the beamsplitter approaches 98.6\% efficiency---meaning only 1.4\% of the population is lost to parasitic states when at the optimal pulse area of $\pi/2$. Similarly, the mirror pulse shown in \Fig \ref{fig:2DAtomOptics-DD-RabiOsc}(c) exhibits an efficiency of 98\% at a pulse area of $\pi$. However, as $\Omega_{\rm Rabi}$ increases, the efficiency of the double-diffraction atom optics decreases. This is primarily a result of the non-negligible coupling to parasitic states due to the finite bandwidth the atom optics pulse, which increases in proportion to $\Omega_{\rm Rabi}$. For instance, when $\Omega_{\rm Rabi} = 0.5\,\delta_{\mu}^{\rm R}$ as in \Figs \ref{fig:2DAtomOptics-DD-RabiOsc}(b,d), the efficiency reduces to $\sim 63\%$ ($83\%$) for the beamsplitter (mirror) and steadily declines as the Rabi frequency increases---where one preferentially populates $\ket{1, \pm\hbar k_x, \pm\hbar k_y}$ and $\ket{1, \pm\hbar k_x, \mp\hbar k_y}$ over the target states.

Although these zero-velocity simulations give a simple picture of the double-diffraction process, the full state dynamics are velocity-dependent and an accurate physical description of these atom optics must include the velocity distribution of the sample. Such a study is beyond the scope of this work, but will be the subject of a future publication.

\section{Derivation of the Multi-Dimensional Mach-Zehnder Phase Shift}
\label{sec:ABDCMatrix}

Here, we give a brief review of the ABCD$\xi$ matrix approach and the general formulas for the AI phase shift based on the theoretical work of Bord\'{e} and Antoine \cite{Borde2001, Antoine2003a, Antoine2003b, Borde2004}. The ABCD$\xi$ formalism for atom optics, based on the ABCD matrix method used in classical optics, is a powerful approach for computing the trajectories and corresponding phase shift for any matter-wave interferometer configuration. The ABCD$\xi$ theorem \cite{Borde2001} gives the exact solution for a general wavepacket evolving in a time-dependent external Hamiltonian that is at most quadratic in position and momentum. Briefly, from the mid-point theorem, the interferometer phase shift can be written as
\be
  \label{DeltaPhi-2}
  \Delta\Phi = \sum_{i=1}^N (\bm{k}_{{\rm A},i} - \bm{k}_{{\rm B},i}) \cdot \bm{Q}_i + \phi_{{\rm A},i} - \phi_{{\rm B},i},
\ee
where $\bm{k}_{{\rm A},i}$ and $\bm{k}_{{\rm B},i}$ are effective wavevectors corresponding to the photon momentum transferred to the atom along paths A and B, respectively, at interaction time $t = t_i$. The vector $\bm{Q}_i$ is the position on the mid-point trajectory
\be
  \bm{Q}_i \equiv \half\big(\bm{q}_{\rm A}(t_i) + \bm{q}_{\rm B}(t_i)\big),
\ee
and the phases $\phi_{{\rm A},i}$ and $\phi_{{\rm B},i}$ are relative laser phases. Equation \eqref{DeltaPhi-2} states that the interferometer phase shift is solely determined by the phase imprinted on the atoms by the lasers at locations along the mid-point trajectory during interaction times $t_1, t_2, \ldots, t_N$. This relationship was recently tested experimentally to a high degree of accuracy \cite{Asenbaum2017}.

The atomic position and momentum trajectories, denoted by the three-component vectors $\bm{q}$ and $\bm{p}$, are computed from the solution to Hamilton's equation \cite{Antoine2003a}
\be
  \label{chij}
  \bm{\chi}_j =
  \begin{pmatrix}
    \bm{q}(t_j) \\
    \bm{p}(t_j)/m
  \end{pmatrix}
  = \mathcal{M}_{ij} \bm{\chi}_i + \bm{\mathcal{L}}_{ij},
  \;\;\;\;\;
  \mathcal{M}_{ij}
  = \begin{pmatrix}
    A_{ij} & B_{ij} \\
    C_{ij} & D_{ij}
  \end{pmatrix},
  \;\;\;\;\;
  \bm{\mathcal{L}}_{ij}
  = \begin{pmatrix}
    \bm{\xi}_{ij} \\
    \bm{\eta}_{ij}
  \end{pmatrix}.
\ee
where the $6 \times 6$ ABCD matrix $\mathcal{M}_{ij} \equiv \mathcal{M}(t_i,t_j)$ and the six-component vector $\bm{\mathcal{L}}_{ij} \equiv \bm{\mathcal{L}}(t_i,t_j)$ are evaluated over the time interval $t \in [t_i,t_j]$. The $3 \times 3$ matrices $A,B,C,D$, and the three-component vectors $\bm{\xi}, \bm{\eta}$, depend on the exact form of the Hamiltonian. The mid-point trajectory at the time of each light pulse can be computed recursively from
\be
  \label{chiRecursion}
  \bm{\chi}_{i+1} = \mathcal{M}_{i,i+1} \left[\bm{\chi}_i +
  \begin{pmatrix}
    \bm{0} \\
    \frac{\hbar}{m} \bm{k}_i
  \end{pmatrix} \right] + \bm{\mathcal{L}}_{i,i+1},
\ee
with initial conditions $\bm{\chi}_0 = (\bm{q}_0, \bm{p}_0/m)$ and $\bm{k}_0 = \bm{0}$.

To give a concrete example of the ABCD$\xi$ formalism, we consider the following Hamiltonian which includes the effects of rotations, accelerations, and gravity gradients---encompassing virtually all of the important physics for atom interferometry applications
\be
  \label{HextSimp}
  \mathcal{H}_{\rm ext} = \frac{\bm{p}^2}{2m} - \bm{p} \cdot (\bm{\Omega} \times \bm{q}) - \frac{m}{2} \bm{q} \cdot \gamma \cdot \bm{q} - m \bm{a} \cdot \bm{q}.
\ee
Hamilton's equation can then be written compactly as
\be
  \label{Hamilton}
  \frac{\dd\bm{\chi}}{\dd t}
  = \begin{pmatrix} \nabla_{\bm p} \mathcal{H}_{\rm ext} \\ -\frac{1}{m} \nabla_{\bm q} \mathcal{H}_{\rm ext} \end{pmatrix}
  = \begin{pmatrix}
    \alpha & \beta \\
    \gamma & \delta \end{pmatrix} \bm{\chi}(t)
  + \begin{pmatrix} \bm{e} \\ \bm{f} \end{pmatrix},
\ee
where the matrices $\alpha$, $\beta$, $\gamma$, $\delta$, and the vectors $\bm{e}$ and $\bm{f}$ are given by
\be
  \label{GammaElements}
  \alpha = \begin{pmatrix}
    0 & \Omega_z & -\Omega_y \\
    -\Omega_z & 0 & \Omega_x \\
    \Omega_y & -\Omega_x & 0
  \end{pmatrix} \!= \delta, \;\;\;
  \beta = \begin{pmatrix}
    1 & 0 & 0 \\
    0 & 1 & 0 \\
    0 & 0 & 1
  \end{pmatrix}\!, \;\;\;
  \gamma = \begin{pmatrix}
    \gamma_{xx} & \gamma_{xy} & \gamma_{xz} \\
    \gamma_{yx} & \gamma_{yy} & \gamma_{yz} \\
    \gamma_{zx} & \gamma_{zy} & \gamma_{zz}
  \end{pmatrix}\!, \;\;\;
  \bm{e} = \bm{0}, \;\;\;
  \bm{f} = \bm{a}.
\ee
Here, $\alpha$ and $\delta$ are the skew-symmetric matrix representation of the rotation vector $\bm{\Omega} = (\Omega_x, \Omega_y, \Omega_z)$, and $\gamma$ contains the elements of the gravity-gradient tensor. In the absence of relativistic effects such as gravitational waves, $\beta = I$ the $3 \times 3$ identity matrix \cite{Borde2004}. Finally, the vector $\bm{f}$ contains the sum of external and gravitational acceleration vectors $\bm{a} = (a_x, a_y, a_z)$.

For the time-independent Hamiltonian \eqref{HextSimp}, the ABCD matrix $\mathcal{M}$ and the vector $\bm{\mathcal{L}}$ can be obtained by directly integrating \Eq \eqref{Hamilton}
\begin{subequations}
\begin{align}
  \mathcal{M}(t_0,t)
  & = \exp\left[ \begin{pmatrix}
    \alpha & \beta \\
    \gamma & \delta \end{pmatrix} (t - t_0) \right], \\
  \bm{\mathcal{L}}(t_0,t)
  & = \int_{t_0}^t \mathcal{M}(t_0,t') \! \begin{pmatrix} \bm{e} \\ \bm{f} \\ \end{pmatrix} \dd t'
   = \begin{pmatrix}
    \alpha & \beta \\
    \gamma & \delta \end{pmatrix}^{\!\! -1} \! \left( 1 - \exp\left[ \begin{pmatrix}
    \alpha & \beta \\
    \gamma & \delta \end{pmatrix} (t - t_0) \right] \right) \! \begin{pmatrix} \bm{e} \\ \bm{f} \\ \end{pmatrix}.
\end{align}
\end{subequations}
We then compute these quantities up to fourth-order in time, and to lowest order in the components of rotation, acceleration, and gravity gradients:
\begin{subequations}
\label{ABCDxieta}
\begin{align}
  A_{ij} & = I + \alpha t_{ij} + \frac{1}{2} \gamma t_{ij}^2, \\
  B_{ij} & = I t_{ij} + \alpha t_{ij}^2 + \frac{1}{6} \gamma t_{ij}^3, \\
  C_{ij} & = \gamma t_{ij}, \\
  D_{ij} & = A_{ij}, \\
  \bm{\xi}_{ij} & = \frac{1}{2} \bm{a} t_{ij}^2 + \frac{1}{3} \alpha \cdot \bm{a} t_{ij}^3 + \frac{1}{24} \gamma \cdot \bm{a} t_{ij}^4, \\
  \bm{\eta}_{ij} & = \bm{a} t_{ij} + \frac{1}{2} \alpha \cdot \bm{a} t_{ij}^2 + \frac{1}{6} \gamma \cdot \bm{a} t_{ij}^3.
\end{align}
\end{subequations}
Here, $t_{ij} = t_j - t_i$ is the time between the $i^{\rm th}$ and $j^{\rm th}$ light pulse, and we note that $\bm{\eta} = \dot{\bm{\xi}} - \alpha\cdot\bm{\xi}$ (\ie $\bm{v} = \dot{\bm{q}} - \alpha\!\cdot\!\bm{q}$) since we consider the atomic motion in a rotating frame. For a Mach-Zehnder geometry, the three pulses are equally spaced in time ($t_{23} = t_{12} = T$). This means that $A_{23} = A_{12}$, $B_{23} = B_{12}$, $\ldots$, and one can write a recursive expression for the positions $\bm{q}_i$ and velocities $\bm{v}_i$ by inserting \Eqs \eqref{ABCDxieta} into \eqref{chiRecursion}. Hence, for pulses $i = 1,2$ one can show
\begin{subequations}
\label{qvi}
\begin{align}
  \bm{q}_{i+1}
  & = \bm{q}_i + \left(\! \bm{v}_i + \frac{\hbar}{m}\bm{k}_i + \alpha\!\cdot\!\bm{q}_i \!\right) T + \frac{1}{2} \left[ \bm{a} + 2\alpha\!\cdot\!\left(\! \bm{v}_i + \frac{\hbar}{m}\bm{k}_i \!\right) + \gamma\!\cdot\!\bm{q}_i \right] T^2 \\
  & + \frac{1}{6} \left[ 2\alpha\!\cdot\!\bm{a} + \gamma\!\cdot\!\left(\! \bm{v}_i + \frac{\hbar}{m}\bm{k}_i \!\right)\right] T^3 + \frac{1}{24} \gamma\!\cdot\!\bm{a} T^4, \nonumber \\
  \bm{v}_{i+1}
  & = \bm{v}_i + \frac{\hbar}{m}\bm{k}_i + \left[\bm{a} + \alpha\!\cdot\!\left(\! \bm{v}_i + \frac{\hbar}{m}\bm{k}_i \!\right) + \gamma\!\cdot\!\bm{q}_i\right] T + \frac{1}{2} \left[ \alpha\!\cdot\!\bm{a} + \gamma\!\cdot\!\left(\! \bm{v}_i + \frac{\hbar}{m}\bm{k}_i \!\right) \right] T^2 + \frac{1}{6} \gamma\!\cdot\!\bm{a} T^3.
\end{align}
\end{subequations}
The mid-point position $\bm{Q}_i$ is the average of the positions $\bm{q}_{{\rm A},i}$ and $\bm{q}_{{\rm B},i}$, which are obtained from \Eqs \eqref{qvi} by replacing $\bm{k}_i$ with $\bm{k}_{{\rm A},i}$ and $\bm{k}_{{\rm B},i}$ for paths A and B, respectively. From the mid-point theorem \eqref{DeltaPhi-2}, the Mach-Zehnder phase shift can be written in the simple form
\be
  \label{DeltaPhi-3}
  \Delta\Phi_{\rm MZ} = \Delta\bm{K}_1\!\cdot\!\bm{Q}_1 + \Delta\bm{K}_2\!\cdot\!\bm{Q}_2 + \Delta\bm{K}_3\!\cdot\!\bm{Q}_3,
\ee
where $\Delta\bm{K}_i \equiv \bm{k}_{{\rm A},i} - \bm{k}_{{\rm B},i}$ and we have omitted the laser phases $\phi_{{\rm A},i}$ and $\phi_{{\rm B},i}$ for clarity. Inserting \Eqs \eqref{qvi} into \Eq \eqref{DeltaPhi-3}, one can show
\begin{align}
\begin{split}
  \label{PhiMZ}
  \Delta\Phi_{\rm MZ} & = \left( \Delta\bm{K}_1 + \Delta\bm{K}_2 + \Delta\bm{K}_3 \right)\!\cdot\!\bm{q}_1 + \left[ \left( \Delta\bm{K}_2 + 2\Delta\bm{K}_3 \right)\!\cdot\!\left(\! \bm{v}_1 + \frac{\hbar}{m}\bm{K}_1 + \alpha\!\cdot\!\bm{q}_1 \!\right) + \Delta\bm{K}_3\!\cdot\!\frac{\hbar}{m}\bm{K}_2 \right] T \\
  & + \frac{1}{2} \left\{ \left( \Delta\bm{K}_2 + 4\Delta\bm{K}_3 \right) \!\cdot\! \left[ \bm{a} + 2\alpha\!\cdot\! \left(\! \bm{v}_1 + \frac{\hbar}{m}\bm{K}_1 \!\right) + \gamma\!\cdot\!\bm{q}_1 \right] +  \Delta\bm{K}_3\!\cdot\!\alpha\!\cdot\!\frac{\hbar}{m}\bm{K}_2 \right\} T^2 \\
  & + \frac{1}{6} \left\{ \left( \Delta\bm{K}_2 + 8\Delta\bm{K}_3 \right)\!\cdot\! \left[ 2\alpha\!\cdot\!\bm{a} + \gamma\!\cdot\! \left(\! \bm{v}_1 + \frac{\hbar}{m}\bm{K}_1 \!\right) \right] + \Delta\bm{K}_3\!\cdot\!\gamma\!\cdot\! \frac{\hbar}{m}\bm{K}_2 + 12\Delta\bm{K}_3 \!\cdot\! \alpha \!\cdot\! \gamma \!\cdot\! \bm{q}_1 \right\} T^3 \\
  & + \frac{1}{24} \left[ \left( \Delta\bm{K}_2 + 16\Delta\bm{K}_3 \right) \!\cdot\! \gamma \!\cdot\! \bm{a} + 32 \Delta\bm{K}_3 \!\cdot\! \alpha \!\cdot\! \gamma \!\cdot\! \left(\! \bm{v}_1 + \frac{\hbar}{m}\bm{K}_1 \!\right) \right] T^4,
\end{split}
\end{align}
where $\bm{K}_i \equiv \frac{1}{2}(\bm{k}_{{\rm A},i} + \bm{k}_{{\rm B},i})$ is the mean momentum transfer wavevector. The first term in \Eq \eqref{PhiMZ} is a phase-matching condition which, in all cases, must vanish in order to obtain spatially-independent interference---implying that $\Delta\bm{K}_1 + \Delta\bm{K}_2 + \Delta\bm{K}_3 = \bm{0}$. This conditions yields perfect overlap of the two interferometer arms at the final beamsplitter. The second term in \Eq \eqref{PhiMZ} is a Doppler phase proportional to $T$. This phase is eliminated in 1D Mach-Zehnder interferometers by using symmetrically-timed pulses (\ie $t_{12} = t_{23} = T$) \cite{Dubetsky2006}. In general, this term is present in the multi-dimensional case because we consider non-parallel wavevectors $\Delta\bm{K}_i$. However, it again vanishes for symmetric AI geometries where ($i$) the pulse separations are identical, ($ii$) the wavevectors during the mirror pulse obey $\bm{k}_{\rm A,2} + \bm{k}_{\rm B,2} = \bm{0}$, and ($iii$) the two beamsplitters satisfy $\bm{k}_{\rm A,3} - \bm{k}_{\rm B,3} = \bm{k}_{\rm A,1} - \bm{k}_{\rm B,1}$ in order to close the interferometer pathways. These three wavevector conditions can be summarized as follows
\be
  \label{KConditions}
  \bm{K}_2 = \bm{0}, \;\;\;
  \Delta\bm{K}_3 = \Delta\bm{K}_1, \;\;\;
  \Delta\bm{K}_2 = -2\Delta\bm{K}_1.
\ee
Inserting these relations into \Eq \eqref{PhiMZ}, the expression for the Mach-Zehnder phase shift simplifies significantly
\begin{align}
\begin{split}
  \label{PhiMZ-2}
  \Delta\Phi_{\rm MZ}
  & = \Delta\bm{K}_1 \!\cdot\! \left[ \bm{a} + \gamma\!\cdot\!\bm{q}_1 + 2\alpha\!\cdot\! \left(\! \bm{v}_1 + \frac{\hbar}{m}\bm{K}_1 \!\right) \right] T^2
  + \Delta\bm{K}_1 \!\cdot\! \left[ 2\alpha\!\cdot\! \big( \bm{a} + \gamma \!\cdot\! \bm{q}_1 \big) + \gamma\!\cdot\! \left(\! \bm{v}_1 + \frac{\hbar}{m}\bm{K}_1 \!\right) \right] T^3 \\
  & + \Delta\bm{K}_1 \!\cdot\! \left[ \frac{7}{12} \gamma \!\cdot\! \bm{a} + \frac{4}{3} \alpha \!\cdot\! \gamma \!\cdot\! \left(\! \bm{v}_1 + \frac{\hbar}{m}\bm{K}_1 \!\right) \right] T^4.
\end{split}
\end{align}
To leading order in $T$, and in the absence of gravity gradients, this expression reduces to \Eq (5) in the main text. This result is completely general in the sense that the wavevector pairs $\bm{k}_{{\rm A},i}$ and $\bm{k}_{{\rm B},i}$ can be oriented in any direction subject to conditions \eqref{KConditions}, and are not necessarily parallel. In contrast to 1D geometries (where $\Delta\bm{K}_1$ and $\bm{K}_1$ are parallel), here we have $\Delta\bm{K}_1\times\bm{K}_1 \neq 0$---hence terms containing the products $\Delta\bm{K}_1\!\cdot\!\alpha\!\cdot\!\bm{K}_1$ and $\Delta\bm{K}_1\!\cdot\!\gamma\!\cdot\!\bm{K}_1$ in \Eq \eqref{PhiMZ-2} are non-vanishing. This yields additional sensitivity to rotations and gravity gradients which has not yet been exploited experimentally.

This treatment can easily be extended to other geometries involving multiple light pulses in one or more dimensions \cite{Dubetsky2006, Cadoret2016} or to large momentum transfer atom optics \cite{Malinovsky2003, Clade2009, Kovachy2012, McDonald2013, Kotru2015, Jaffe2018}.

\end{document}